\documentclass[12pt,nohyper,letterpaper]{JHEP3}         
\usepackage{amssymb,amsfonts}

\usepackage{epsfig}
\usepackage{graphicx}

\def\be{\begin{eqnarray}}
\def\ee{\end{eqnarray}}
\newcommand{\nn}{\nonumber}
\newcommand\para{\paragraph{}}
\newcommand{\ft}[2]{{\textstyle\frac{#1}{#2}}}
\newcommand{\eqn}[1]{(\ref{#1})}

\def\Dslash{\,\,{\raise.15ex\hbox{/}\mkern-12mu D}}
\def\Dbarslash{\,\,{\raise.15ex\hbox{/}\mkern-12mu {\bar D}}}
\def\delslash{\,\,{\raise.15ex\hbox{/}\mkern-9mu \partial}}
\def\delbarslash{\,\,{\raise.15ex\hbox{/}\mkern-9mu {\bar\partial}}}
\def\pslash{\,\,{\raise.15ex\hbox{/}\mkern-9mu p}}
\def\calDslash{\,\,{\raise.15ex\hbox{/}\mkern-12mu {\cal D}}}

\newcommand{\pu}{{+\uparrow}}
\newcommand{\pd}{+\downarrow}
\newcommand{\minu}{-\uparrow}
\newcommand{\mind}{-\downarrow}

\def\lae{\mathrel{\mathop{\smash{\lower .5 ex \hbox{$\stackrel<\sim$}}}}}
\def\lae{\mathrel{\mathop{\smash{\lower .5 ex \hbox{$\stackrel>\sim$}}}}}

\title{A Holographic Flat Band}
\author{Jo\~ao N. Laia and David Tong
 \\
Department of Applied Mathematics and Theoretical Physics, \\
University of Cambridge, UK\\
{\tt j.laia, d.tong@damtp.cam.ac.uk}
}

\abstract{We describe a novel implementation of non-relativistic fermions in AdS/CFT by imposing Lorentz violating boundary terms for a Dirac spinor in AdS${}_4$.  The dual boundary theory is scale invariant and exhibits a number of interesting properties, including a dispersionless flat band of gapless excitations.}

\begin{document}
\pagestyle{plain} \setcounter{page}{1}
\newcounter{bean}
\baselineskip16pt

\section{Introduction}

A {\it flat band} is a  dispersion relation that does not disperse. The kinetic energy of an excitation is suppressed, resulting in  a Hamiltonian that admits a large degeneracy of localised, single particle eigenstates. The simplest examples of flat bands are very familiar: Landau levels. However, in the absence of a magnetic field, they are more exotic beasts. Nonetheless, flat bands can be manufactured in lattice models, typically by arranging a form of geometric frustration in which different hopping terms interfere.

\para
When flat bands are partially filled, interesting things happen. Arbitrarily small interactions or perturbations will lift the degeneracy, often driving the system to a highly correlated ground state. This mechanism provides 
the springboard for the rich spectrum of quantum Hall phases emerging from Landau levels. Moreover, in various lattice models, the flat band degeneracy has been used to  drive instabilities towards itinerant ferromagnetism
\cite{miekle,tasaki} and superconductivity \cite{imada}, as well as to fractional quantum Hall states  \cite{katsura,green,wen,sun}. Flat bands have also made an appearance in studies of optical lattices \cite{wu}, bilayer graphene \cite{graphene} and topological phases  \cite{volovik}.



\para
The purpose of this paper is to describe the emergence of a flat, dispersionless spectrum for fermions in a holographic framework.
Of course, in the context of AdS/CFT, highly correlated ground states are our bread and butter and we have little need to build them using flat bands. Nonetheless, we hope that the construction of a holographic flat band will allow a further, controlled study of this phenomenon.

\para
Concretely, we construct a non-relativistic boundary theory in $d=2+1$ dimensions. However, in contrast to other gauge/gravity implementations of non-relativistic field theories, we do not change the bulk geometry. Instead we work in a bulk AdS${}_4$ spacetime and break Lorentz invariance only through the boundary conditions imposed on a Dirac spinor field. As we will see, it is possible to impose boundary conditions which preserve rotational invariance and scale invariance, but break both boosts and parity. The resulting 3d boundary theory can be thought of as a relativistic CFT, described by the ``$N^{2\,}$" degrees of freedom of the bulk geometry, coupled to a 3d Dirac spinor in a manner that breaks Lorentz invariance but, at least to leading order in $1/N$, preserves $z=1$ scale invariance.

\para
We shall see that the effect of the Lorentz breaking boundary conditions is rather brutal. The speed of propagation of the boundary fermion is driven to zero, resulting in an infinite flat band, comprising of dispersionless, gapless excitations for all values of the spatial momenta.

\para
In Section 2 we describe the novel boundary conditions, together with some features of the dual boundary field theory. Notably, the theory has a marginal, double trace operator, independent of the bulk fermion mass $m$. In some sense, this marginal operator can be thought of as the Goldstone mode for the broken Lorentz symmetry.
In Section 3 we describe the single particle Green's function in vacuum and demonstrate the existence of the flat band.

\para
In Section 4, we place our theory at finite temperature and chemical potential by replacing the bulk with an AdS black hole. There has been much recent activity in the study of spinor two-point functions in these backgrounds
due to the emergence of a Fermi surface revealing a rich spectrum of non-Fermi liquid behaviour \cite{ssl,mit1,leiden,mit2}. We find these Fermi surfaces numerically with our novel boundary conditions. We will see that, perhaps surprisingly, despite our singular spectrum, the formation and excitations of the Fermi surface are not greatly changed from the Lorentz invariant theory.



\section{Non-Relativistic Fixed Points}

Gravity duals for non-relativistic field theories are typically implemented through the construction of a
dual geometry that differs from AdS. The familiar examples are the Lifshitz geometry \cite{lifshitz} and  Schr\"odinger geometry \cite{son,mb}, both of which describe quantum critical points with dynamical exponents $z\neq 1$. In this section we describe a non-relativistic system with $z=1$ scaling symmetry in which Lorentz symmetry is broken by boundary conditions.

\para
To this end, we study a four-component Dirac fermion in asymptotically AdS${}_4$ backgrounds. We start in this section by considering  vacuum
$AdS$ space of unit radius with Lorentzian metric
\be ds^2 = \frac{dr^2}{r^2} + r^2\eta_{\mu\nu}dx^\mu dx^\nu
\nn\ee
A detailed description of fermions in AdS spaces was presented in \cite{iqballiu} whose conventions we (mostly) follow.
The dynamics of a Dirac spinor is governed by the bulk action
\be S_{\rm bulk} =\int d^4x \sqrt{-g}\,i\bar{\psi} \,\left[\frac{1}{2}\left(\Gamma^M\!\stackrel{\rightarrow}D_M-
\stackrel{\leftarrow}D_M\!\Gamma^M\right)-m\right]\psi\label{bulkaction}\ee
The covariant derivative is $D_M = \partial_M + \ft14 \omega_{ab,M}\Gamma^{ab}$ where $\omega_{ab,M}$ is the bulk spin connection and $\Gamma^{ab} = \ft12 [\Gamma^a,\Gamma^b]$. In all these expressions, the capital index $M$ denotes bulk spacetime while $a,b$ denote bulk tangent space.
The action \eqn{bulkaction} is not complete without boundary terms and we will spend some time shortly detailing the possibilities.

\para
To write the Dirac equation, it is convenient to choose the basis of bulk gamma matrices,
\be \Gamma^\mu = \left(\begin{array}{cc} 0 & \gamma^\mu \\ \gamma^\mu & 0 \end{array}\right),\  \mu =0,1,2\ \ \ ,\ \ \ \
\Gamma^r = \left(\begin{array}{cc} 1 & 0 \\ 0 & -1 \end{array}\right) \nn\ee
where $\gamma^\mu$ furnish a representation of the 3d Clifford algebra. We will later choose the basis $\gamma^\mu = (i\sigma^3, \sigma^1,\sigma^2)$. The Dirac equation can then be expressed in terms two-component spinors $\psi_+$ and $\psi_-$,  each of which is an eigenvector  of $\Gamma^r$.
\be \psi = \left(\begin{array}{c} \psi_+ \\ \psi_-\end{array}\right)\nn\ee
Working in Fourier modes for the boundary directions, $\psi(x,r) = \psi(k;r)e^{ik\cdot x}$, the bulk Dirac equation is
\be r\left(r\partial_r +\frac{3}{2}-m\right)\psi_+ &=& -{ik\cdot\gamma}\, \psi_-\nn\\ r\left(r\partial r +\frac{3}{2}+m\right)\psi_- &=&  {ik\cdot \gamma} \,\psi_+\nn\ee
Before discussing boundary conditions, it is useful to first examine the behaviour of the solutions near the AdS boundary $r\rightarrow0$ where, to leading order,
\be
 \psi_+ (k;r)&=& A(k) r^{-3/2+m}+B(k) r^{-3/2 -m -1} + \ldots
 \nn\\ \psi_- (k;r)&=& D(k)r^{-3/2-m} + C(k)r^{-3/2+m-1} + \ldots \ \ \ \ \ \ {\rm as}\ r\rightarrow \infty
\label{asym}\ee
Here the 2-component spinors $A$, $B$, $C$ and $D$ are related by
\be
 D=- \frac{i \gamma . k}{k^2} (2m +1) B \qquad, \qquad C= \frac{i \gamma . k}{2m-1} A
\label{relate}\ee
We must decide which of these spinors is interpreted as the source and which is the response. As we now review, this is determined by the choice of boundary terms.

\subsection{Boundary Terms}

The bulk action \eqn{bulkaction} alone does not provide a full description of the dynamics. The problem lies with the terms which arise after integrating by parts. These are evaluated on the boundary of AdS
\be \delta S_{\rm bulk} = \frac{i}{2}\int_{\partial {\cal M}} d^3x\sqrt{-h} \left(\delta\bar{\psi}_+\psi_--\delta\bar{\psi}_-\psi_+ -\bar{\psi}_+\delta\psi_- + \bar{\psi}_-\delta\psi_+\right) + \ \mbox{bulk term}
\nn\ee
Here $h$ is the determinant of the induced boundary metric, $h=gg^{rr}$.

\para
Because the Dirac action is first order, it is not acceptable to insist that both $\delta\psi_-$ and $\delta \psi_+$ vanish on the boundary. Restricting all components of the spinor in this way  would be akin to fixing both position and momentum \cite{henneaux}. Instead we must  fix just half of the spinor components which usually means either $\psi_+$ or $\psi_-$ although we shall discuss other options below. To achieve this, the bulk action must be augmented by a boundary term so that together they give rise to a well-defined variational principle. Depending on the value of the bulk mass $m$, there are a number of different ways to achieve this.

\subsubsection*{Standard Quantization}

For  $m>0$, the dominant term in \eqn{asym} as we approach the boundary is  $A(k)$. In the usual manner of holography, we would expect to fix the coefficient $A(k)$, treating it as a  source for the dual operator $\Psi$ in the boundary theory.  This is achieved by adding the boundary action
\be S_{\rm bdy} = \frac{i}{2}\int_{\partial {\cal M}} d^3x\sqrt{-h}\,\bar{\psi}\psi = \frac{i}{2}\int_{\partial {\cal M}} d^3x\sqrt{-h}\left(\bar{\psi}_-\psi_+ +\bar{\psi}_+\psi_-\right)
\label{stbc}\ee
The variation of the full action is then
\be \delta S_{\rm bulk} + \delta S_{\rm bdy} = i\int_{\partial {\cal M}} d^3x\sqrt{-h}\left(  \delta\bar{\psi}_+\psi_- +\bar{\psi}_-\delta \psi_+ \right)
\nn\ee
which indeed vanishes if we impose Dirichlet boundary conditions for  $\psi_+$. This choice of boundary condition is usually referred to as the {\it standard quantization} for fermions. The dual boundary field theory is a Lorentz invariant conformal field theory and contains a 3d Dirac fermion operator $\Psi$, with dimension
\be \Delta_+ = \frac{3}{2}+m\nn\ee

\subsubsection*{Alternative Quantization}

While the $A(k)$ term in \eqn{asym} is always dominant for $m>0$, something rather special happens in the window $0\leq m < 1/2$. Here the fall-off of $A(k)$ becomes normalizable \cite{iqballiu}. This means that while we are still at liberty to fix $\psi_+$ on the boundary if we want to, we are no longer obliged to do so since $\psi_+$ will happily fluctuate if we allow it. This provides us with an alternative choice of boundary condition in which we instead fix $\psi_-$ on the boundary. It is a simple matter to implement this: we need only add the boundary term with opposite sign,
\be S_{\rm bdy} = -\frac{i}{2}\int_{\partial {\cal M}} d^3x\sqrt{-h}\,\bar{\psi}\psi 
\label{altbc}\ee
which now results in a well defined variational principle if we impose Dirichlet conditions  for $\psi_-$. This is known as {\it alternative quantization}.

\para
The ambiguity of boundary conditions for fields in AdS that lie within a particular mass window  was first recognised in \cite{bf} and its implication for the AdS/CFT correspondence was explained in  \cite{kw}: the two different boundary conditions correspond to two different dual CFTs on the boundary. In this alternative CFT, the  fermion operator $\Psi$ has dimension
\be \Delta_- = \frac{3}{2} - m\label{altdim}\ee
Correlation functions in this alternative CFT are related to those in the standard quantization by a Legendre transform \cite{kw}. 
Note that as  $m\rightarrow 1/2$, the dimension of the operator tends towards the unitarity bound $\Delta_- \rightarrow 1$. 
Finally, note that taking the mass $m<0$ simply exchanges the role of the standard and alternative quantizations.

\subsubsection*{Mixed Quantization}

If we are interested in Lorentz invariant boundary CFTs, then the two boundary conditions \eqn{stbc} and \eqn{altbc} exhaust the possibilities. Here we would like to relax this condition. Specifically, we will look for boundary conditions which preserve both  the $U(1)$ global symmetry $\psi \rightarrow e^{i\theta}\psi$, rotational invariance and scale invariance, but not the full Lorentz invariance.

\para
There are two such choices, related by a discrete symmetry. The first can be implemented by the boundary condition
\be S_{\rm bdy} = \frac{1}{2}\int_{\partial {\cal M}}d^3x\sqrt{-h} \ \bar{\psi}\,\Gamma^1\Gamma^2\psi =
-\frac{1}{2}\int_{\partial {\cal M}}d^3x\sqrt{-h} \left({\psi}^\dagger_-\psi_+ +{\psi}^\dagger_+\psi_-\right)
\label{mixedbc}\ee
With the basis of 3d gamma matrices $\gamma^\mu = (i\sigma^3, \sigma^1,\sigma^2)$, we write each two component spinor as
\be \psi_+ =\left(\begin{array}{c}\psi_{\pu} \\ \psi_{\pd}\end{array}\right)\ \ \ ,\ \ \  \psi_- =\left(\begin{array}{c}\psi_{\minu} \\ \psi_{\mind}\end{array}\right)\ee
The variation of the full action is now
\be  \delta S_{\rm bulk} + \delta S_{\rm bdy} = -\int_{\partial {\cal M}} d^3x\sqrt{-h}\left(  \delta\psi_{\pu}^\dagger\psi_{\minu} + \psi_{\minu}^\dagger\delta\psi_{\pu} + \psi_{\pd}^\dagger \delta\psi_{\mind}+\delta\psi_{\mind}^\dagger\psi_{\pd}\right)\nn\ee
This vanishes if we impose Dirichlet conditions for $\psi_{\pu}$ and $\psi_{\mind}$.

\para
The boundary field theory once again contains two fermionic operators, $\Psi_\uparrow$ and $\Psi_\downarrow$.
The 3d rotation operator is $R=-\ft{i}{4}[\gamma^1,\gamma^2] = \ft12\sigma^3$ which means  that $\Psi_{\uparrow}$ has spin $+\ft12$ and $\Psi_{\downarrow}$ has spin $-\ft12$. However now the dimensions of these two operators differ, reflecting the broken Lorentz invariance. They are
\be \Delta[\Psi_\uparrow] = \Delta_+\ \ \ ,\ \ \ \Delta[\Psi_\downarrow] = \Delta_-\label{mixeddim}\ee

\subsubsection*{Discrete Symmetries}

There is a second mixed quantization where the boundary condition \eqn{mixedbc} differs by an overall minus sign, requiring that we impose Dirichlet conditions for  $\psi_{\pd}$ and $\psi_{\minu}$. The resulting boundary operators have dimension $\Delta[\Psi_\uparrow] = \Delta_-$ and $\Delta[\Psi_\downarrow] = \Delta_+$.

\para
These two conformal field theories are related by a parity transformation. Recall that in $d=2+1$ dimensions, parity acts by reflecting just a single spatial coordinate, $x^1\rightarrow -x^1$ with the corresponding action on spinors $P: \Psi_\uparrow\rightarrow \Psi_\downarrow$. Thus, while the Lorentz invariant theories preserve parity, it is broken by our non-relativistic boundary conditions\footnote{We note that we could equally well implement these mixed boundary conditions for spinors in Lifshitz geometry. Since there is no Lorentz symmetry of the boundary theory in this case, the only price paid is the breaking  of parity.}.
The non-relativistic boundary theory also breaks charge conjugation, which acts as $C:\Psi_\uparrow \rightarrow \Psi_\downarrow^\star$.

%

\subsection{A Poor Man's RG}

Dual CFTs described by different boundary conditions are not unrelated. There exists a renormalization group (RG) flow from the alternative quantization to the standard quantization, initiated by turning on a relevant  double trace operator \cite{witten,micha}. 

\para
Some properties of this RG flow for fermions were described in \cite{allais}. One starts in the alternative quantization and perturbs the boundary theory by the Lorentz invariant double trace operator $\bar{\Psi}\Psi$. This is a relevant operator, with   dimension $\Delta[\bar{\Psi}\Psi]= 3-2m$. The end point of the flow is  the standard quantization. For this reason, the alternative quantization is sometimes referred to as the UV CFT, while the standard quantization is the IR CFT.

\para
What about our non-relativistic dual CFT? This too can be reached by an RG flow from the alternative quantization.
If we care only about rotational symmetry and the $U(1)$ global symmetry, the UV CFT contains two relevant operators: $\bar{\Psi}\Psi$ and $\bar{\Psi}\gamma^0\Psi$. To flow to the non-relativistic theory with operators \eqn{mixeddim}, we want to change the boundary conditions for the bulk field $\psi_{\pd}$, leaving $\psi_{\pu}$ untouched. This can be achieved by perturbing with the double trace operator
\be \int d^3x\ i\bar{\Psi}(1+i\gamma^0)\Psi= \int d^3x\ 2\Psi^\dagger_{\downarrow}\Psi_\downarrow\label{wheretogo}\ee
It is useful to remind ourselves what these two operators would mean for a free fermion. $\bar{\Psi}\Psi$ is a mass term; it breaks parity, but respects charge conjugation. Meanwhile, $\bar{\Psi}\gamma^0\Psi$ is a chemical potential for a free fermion (at least in a suitable thermodynamic ensemble). It breaks charge conjugation but respects parity. Had we added  the operator \eqn{wheretogo} for a free fermion, it would open up a gap in the spectrum, with the chemical potential tuned to sit at the bottom of the band, resulting in a single component, non-relativistic, gapless fermion. Of course, our fermions are far from free. We will see shortly what becomes of them.

\begin{figure}[t]
   \centering
    \begin{minipage}[b]{10 cm}
    \includegraphics[scale=0.6]{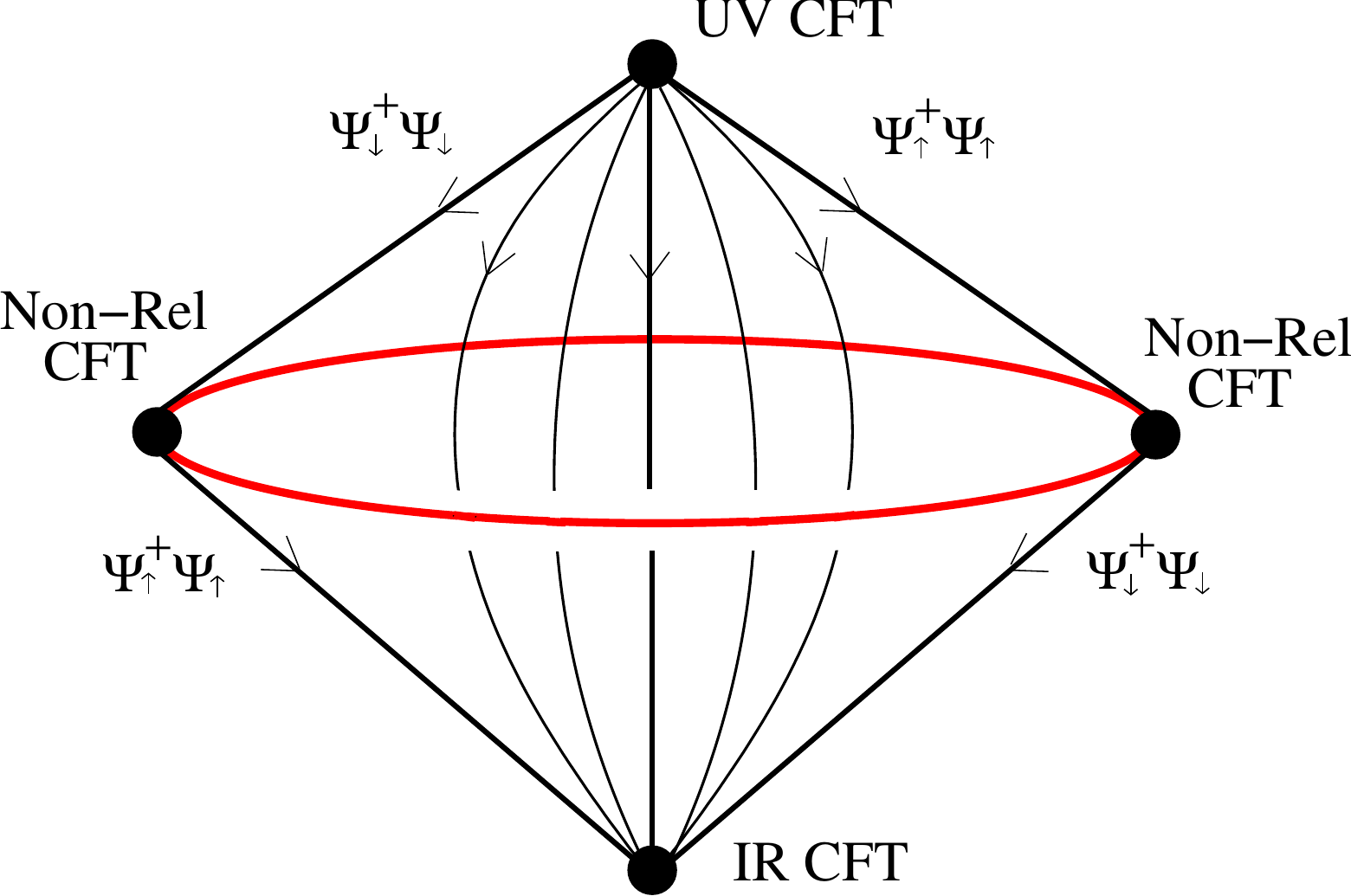}  
  \end{minipage}
  \caption{Flows between the UV, IR and non-relativistic fixed points. The red circle denotes a
${\bf S}^2$ of fixed points, breaking the $U(1)$ global symmetry. Parity reflects the diagram
about the vertical.}
\end{figure}

\para
Perturbing the UV CFT by $\Psi^\dagger_{\uparrow}\Psi_\uparrow$ will result in a flow to the other non-relativistic fixed point. We leave a full analysis of the flow equations to 
future work \cite{fermirg}. For now, it is comforting to know that we can reach the non-relativistic fixed point from a sensible starting point, albeit at the cost of fine tuning an RG flow.

\para
Each of the non-relativistic fixed points also contains a relevant operator,  namely $\Psi_\downarrow^\dagger\Psi_\downarrow$ (or, in the parity related theory, $\Psi_\uparrow^\dagger\Psi_\uparrow$). In each case, turning on this operator takes us back to the IR CFT associated to the standard quantization. The picture of two dimensional RG flows that emerges is shown in Figure 1.

\subsubsection*{A Circle of Fixed Points}

There is one further double trace operator of interest at the non-relativistic fixed point, namely the charged, rotationally invariant operator $\Psi_\uparrow\Psi_\downarrow$. Usually the dimensions of operators are determined by the mass of a bulk field. Not so for this operator: its dimension is $\Delta_++\Delta_-=3$. In other words, it is marginal for all values of the bulk mass in the window $|m|<1/2$.

\para
A similar mechanism for generating a marginal operator in the case of complex scalar fields was described in \cite{witten}: the real part of the scalar was given standard boundary conditions; the imaginary part alternative. In both cases, the marginal operator can be loosely thought of as a Goldstone mode for the symmetries broken by the boundary conditions. For us, the broken symmetry is  Lorentz symmetry.

\para
Turning on the marginal operator $\Psi_\uparrow\Psi_\downarrow$ sweeps out a manifold of fixed points; this is the red circle in Figure 1. (For readers without colour, it is the circle in Figure 1). Since this operator is complex, there is a two-dimensional set of fixed points. The magnitude of the coefficient takes us around the circle; the phase (not shown in the figure) takes us from the back of the circle to the front. The result is that the manifold of fixed points is topologically ${\bf S}^2$. Since we are turning on a charged operator, each of these CFTs away from the two special points breaks the $U(1)$ global symmetry of the boundary theory. In fact, we can define these fixed points directly by introducing the linear combinations,
\be \chi_{\pm\uparrow} = \cos\theta\,\psi_{\pm \uparrow} + \sin\theta\,\psi_{\pm\downarrow}^\dagger\ \ \ ,\ \ \ \chi_{\pm\downarrow} = \cos\theta\,\psi_{\pm \downarrow} - \sin\theta\,\psi_{\pm\uparrow}^\dagger\nn\ee
It is then simple to check that the addition of the boundary term,
\be S_{\rm bdy} = -\frac{1}{2}\int_{\partial {\cal M}}d^3x\sqrt{-h} \left({\chi}^\dagger_-\chi_+ +{\chi}^\dagger_+\chi_-\right)
\label{chargebc}\ee
requires us to impose Dirichlet conditions for $\chi_{+\uparrow}$ and $\chi_{-\downarrow}$. Because $\chi_{+\uparrow}$ is a linear combination of $\psi_{+\uparrow}$ and $\psi^\dagger_{+\downarrow}$, both of which have the same radial dependence, these boundary conditions preserve scale invariance. However, as promised, they break the global symmetry apart from at the 
poles, $\theta=0$ and $\theta=\pi/2$, where these boundary conditions 
reduce to our previous mixed quantization \eqn{mixeddim} and its parity twin.

\para
Finally, the RG properties of a massless bulk fermion, $m=0$, are worthy of comment. 
In this case, there is no RG flow from alternative to standard quantization since both 
$\bar{\Psi}\Psi$ and $\bar{\Psi}\gamma^0\Psi$ are marginal. In particular, the 
marginal operator $\bar{\Psi}\Psi$ gives rise to a circle of fixed points that 
interpolate between the alternative and standard quantization, as previously 
described in \cite{porrati}. This is again a ``Goldstone operator'', associated 
to the breaking of bulk chiral symmetry by the boundary condition. 
Together with the Lorentz violating  
$\bar{\Psi}\gamma^0\Psi$, the two operators sweep out a torus ${\bf T}^2$ 
of charge invariant fixed points. Further including the charge violating 
operator $\Psi_\uparrow \Psi_\downarrow$, results in a four-dimensional 
manifold of fixed points.

\section{Green's Functions}

We now turn to the excitation spectrum of the non-relativistic boundary theory. We will first provide a quick method to determine the Green's function. However, as previously advertised, the resulting spectrum will have the rather peculiar property of a flat band and for this reason we subsequently provide a second derivation.

\para
It will prove useful to first recap the fermion propagator for relativistic conformal theories. The two point function, which  is fixed by symmetries, was first computed holographically in  \cite{henning,mueck}, with various properties of the retarded propagator clarified in \cite{iqballiu}. In this case, the Dirichlet boundary conditions for some components of the spinor at the boundary $r\rightarrow \infty$ are accompanied by ingoing boundary conditions at the AdS horizon $r=0$.  The Dirac equation is then easily solved in terms of appropriate Bessel functions. 

\para
As we reviewed in Section 2, for the standard quantization $\psi_+$ is fixed on the boundary, and the two-spinor $A$ in \eqn{asym} is interpreted as the source, while $D$ is the response. The relation between the two is written 
%
$D(k)=S(k)A(k)$. The retarded
propagator is then defined to be  $G_R(\Delta_+)=-iS\gamma^0$, where the factor of $\gamma^0$ in this expression reflects the fact that the Green's function computes $\langle \Psi^\dagger\Psi\rangle$ rather than $\langle \bar{\Psi}\Psi\rangle$. For timelike momenta, $\omega > |\vec{k}|$, one finds the propagator
\be G_R(\Delta_+) = \alpha_{\Delta_+}\,(\omega^2-|\vec{k}|^2)^{\Delta_+-2}(k\cdot\gamma)\gamma^0\label{relgreen}\ee
where the overall constant is given by
\be \alpha_\Delta =  \frac{e^{-(\Delta-1)\pi i}}{2^{2\Delta-3}}\,\frac{\Gamma(2-\Delta)}{\Gamma(\Delta-1)}\nn\ee
In the alternative quantization, we instead fix $\psi_-$, with $D$ now interpreted as the source and $A$ the response. To compute the propagator, it is not necessary re-solve the Dirac equations in the bulk. We need only realise that the roles of the source and response have been interchanged so that the propagator is given by $G_R(\Delta_-) = iS^{-1}\gamma^0$. Indeed, as a check one can easily verify that the two-point function obeys
\be G_R(\Delta_-)=\gamma^0G_R(\Delta_+)^{-1}\gamma^0\nn\ee

\subsection{Non-Relativistic Boundary Conditions}

With mixed, non-relativistic boundary conditions, the sources are $A_\uparrow$ and $D_\downarrow$. The propagator is again easily determined from \eqn{relgreen} without going through the rigmarole of solving the bulk Dirac equations. We need only massage the equation $D(k)=S(k)A(k)$ into the form,
\be \left(\begin{array}{c}D_\uparrow \\ A_\downarrow\end{array}\right) = {\cal S} \left(\begin{array}{c} A_\uparrow \\ D_\downarrow\end{array}\right)\nn\ee
The retarded propagator is ${\cal G}_R = -{\cal S}$. The lack of Lorentz invariance in the theory means that there is no $\gamma^0$ factor in this relationship and it is simple to check that this definition is consistent with the requirement of unitarity since it ensures that the imaginary part of the diagonal components of the propagator are  positive. For timelike momenta, $\omega > |\vec{k}|$, the retarded propagator is
\be {\cal G}_R = -\frac{1}{\omega}\left(\begin{array}{cc} \alpha_{\Delta_+}(\omega^2-|\vec{k}|^2)^{\Delta_+-1} & ik_1-k_2  \\ -ik_1-k_2 & \alpha_{\Delta_-}(\omega^2-|\vec{k}|^2)^{\Delta_--1} \end{array}\right)\label{gtime}\ee
For $\omega < -|\vec{k}|$, the coefficients $\alpha_{\Delta_\pm}$ in the above expression should be replaced by $\alpha_{\Delta_\pm}^\star$.  It is perhaps unusual to see a two-point function mixing operators of different dimensions but this is not forbidden in a non-relativistic system. (The usual proof that this cannot happen makes crucial use of special conformal transformations). However, by far the most unusual aspect of this propagator is the $1/\omega$ pole out-front. This is the first indication of a flat band. This pole survives at spacelike momenta, $|\vec{k}|^2>\omega^2$, where the retarded Green's function is given by
\be {\cal G}_R = -\frac{1}{\omega+i\epsilon}\left(\begin{array}{cc} |\alpha_{\Delta_+}|(|\vec{k}|^2-\omega^2)^{\Delta_+-1} & ik_1-k_2  \\ -ik_1-k_2 & |\alpha_{\Delta_-}|(|\vec{k}|^2-\omega^2)^{\Delta_--1} \end{array}\right)\label{gspace}\ee
We have added the $+i\epsilon$ prescription to ensure that the propagator is analytic in the upper-half complex $\omega$ plane, as required by causality. We will see later that this is consistent when we  study our system at finite temperature.

\para
To better understand the structure of the propagator, let us look at the eigenvalues, which we will denote as $\lambda_+$ and $\lambda_-$.
The propagator obeys $\det {\cal G}_R=-1$, which tells us that the eigenvalues come in reciprocal pairs. For small $\omega\ll|\vec{k}|$, the two eigenvalues scale as $\lambda_+\sim -1/\omega$ and $\lambda_-\sim\omega$. Specifically,
\be
\lambda_+ \approx -\frac{|\vec{k}|}{\omega}\,\left(|\alpha_{\Delta_+}| \vec{k}^{\,2m} + \frac{1}{|\alpha_{\Delta_+}|\vec{k}^{\,2m}}\right)\ \ \ {\rm and}\ \ \ \lambda_-\approx\frac{\omega}{|\vec{k}|}\,\left(|\alpha_{\Delta_+}| \vec{k}^{\,2m} + \frac{1}{|\alpha_{\Delta_+}|\vec{k}^{\,2m}}\right)^{-1}
\nn\ee
For the special case of $m=0$ the eigenvalues simplify to $\lambda_\pm=\omega^{-1}(\sqrt{\vec{k}^{\,2}-\omega^2}\pm |\vec{k}|)$. The   corresponding eigenvectors are $(|\vec{k}|,\mp (ik_1-k_2))$.

\para
One of the eigenvalues exhibits the peculiar property of a pole at $\omega=0$ {\it for all values} of the spatial momentum $\vec{k}$. This implies the existence of localised, gapless excitations with vanishing speed of propagation. As discussed in the introduction, dispersionless modes of this kind are referred to as a {\it flat band} although, in the absence of a magnetic field, it is more usual to find such a spectrum in  lattice models where the band covers only a finite interval of momenta. The holographic theory exhibits an infinite flat band.

\para
The appearance of a flat band is very surprising. To illustrate how extreme this behaviour is, one only has to note that the very language of physics breaks down: dispersion relations do not disperse; propagators do not propagate. It is a form of local criticality, in which neighbouring spatial points no longer speak, but different from the $z\rightarrow \infty$, AdS${}_2$ criticality that arises at finite density \cite{mit2}. 
Needless to say, it would be interesting to understand the emergence of this flat spectrum from the perspective of the boundary theory. One clue may come from the analysis of a free fermion. As discussed in the previous section, the deformation \eqn{wheretogo} gaps the spectrum of a free fermion and leaves the chemical potential kissing the bottom of the band, resulting a gapless excitation with quadratic dispersion. For our strongly interacting fermion, such a quadratic dispersion is not possible since the coupling to the background AdS spacetime forces the correlation function to enjoy $z=1$ scaling with an associated $\omega = vk$ dispersion. Thus we seem to be in a situation where the deformation \eqn{wheretogo} removes the linear dispersion term but higher order terms are forbidden by the background, driving the speed of propagation $v\rightarrow 0$.

\para
One might worry about the existence of flat bands in a theory which allows for pair creation. If we could populate the band with particle anti-particle pairs from the vacuum, then our theory would be rabidly unstable. Thankfully, that is not the case. The excitations of the band have fixed charge under the global symmetry of the boundary\footnote{Strictly, the boundary theory has a global symmetry only if we gauge the $\psi\rightarrow e^{i\alpha}\psi$ symmetry in the bulk. We implicitly assume this is the case. We will be more explicit in the following section.} and charge conservation requires that any excitation of the flat band with eigenvalue $\lambda_+$ that is created from the vacuum must be accompanied by its partner with eigenvalue $\lambda_-$. Fortunately, this means that the flat band cannot be populated from the vacuum. We must place particles in by hand.

\para
As mentioned in the introduction, a flat band is most interesting because it is unstable to interactions and perturbations, often driving the physics to an interesting strongly coupled corner. In the next section, we will populate and heat our band and find that it is surprisingly stable, at least when the fermions are treated in the probe approximation. The discussion above suggests that the flat band is most  likely a large $N$ artifact and the most interesting question is how, if at all, it is lifted by $1/N$ corrections. One such source of corrections arises from quantum fermion loops in the bulk.  In particular, since the boundary conditions break Lorentz invariance, bulk vacuum bubbles, which usually only renormalize the cosmological constant, are now at liberty to change the bulk metric away from AdS\footnote{We thank D.T. Son for discussions on this point.}. 

\para
Finally, we would like to mention that something similar to a flat band was seen already in holographic electron star calculations, where the Fermi surface is smeared into a Fermi ball, with gapless excitations at all points in the interior \cite{sean1}. This ball is resolved by $1/N$ corrections into a family of discrete nested Fermi surfaces \cite{sean2,semifl}.

\subsection{Bulk Zero Modes}

The derivation of the Green's function presented above used the relationship with the relativistic theory. One can also extract the Green's functions in a more pedestrian manner directly from the bulk solutions. This method has the advantage of exhibiting the normalizable bulk modes corresponding to the $\omega=0$ pole.

\para
The solution to the bulk Dirac equation is well known \cite{mueck}. Restricting to spacelike momenta, $k^2=|\vec{k}|^2-\omega^2>0$, the normalizable solution is given by the Bessel functions
\be \psi_+(r) =  \frac{1}{r^2}\, K_{\nu_+}\left(\frac{k}{r}\right)\chi_+\ \ \ ,\ \ \
\psi_-(r) =  \frac{1}{r^2}\, K_{\nu_-}\left(\frac{k}{r}\right)\chi_- \nn\ee
where $\nu_\pm=\ft12 \pm m$. Expanding to leading order in $r$, the  two-component spinors $\chi_+$ and $\chi_-$ are related to the spinors $A$ and $D$ in \eqn{asym}. However, we must still satisfy \eqn{relate} which provides a relationship between source and response. In this manner, it is a simple matter to reproduce the Green's function \eqn{gspace}.

\para
Poles in the Green's function correspond to normalizable bulk modes in the absence of a source. Thus, the flat band Green's function \eqn{gspace} tells us that the mixed boundary conditions should allow for $\omega=0$ modes for arbitrary $\vec{k}$. Turning off the source  sets $\psi_{+\uparrow} = \psi_{-\downarrow}=0$. The conditions \eqn{relate} then requires that the $\psi_+$ and $\psi_-$ solutions are related by
\be \left(\begin{array}{c} 0 \\ \psi_{+\downarrow}\end{array}\right) \sim \left(\begin{array}{cc} \omega & k_1-ik_2 \\ k_1-ik_2 & -\omega\end{array}\right) \left(\begin{array}{c}\psi_{-\uparrow} \\ 0 \end{array}\right)\nn\ee
When $\omega=0$, this equation can indeed be solved for  arbitrary $\vec{k}$. One can also check that the bulk Hamiltonian, ${\cal H} \sim \bar{\psi}(\gamma^r\partial_r + k_i\gamma^i -m)\psi=0$ evaluated on these solutions, courtesy  of the spinor contractions.

\section{Black Hole Backgrounds}\label{secbh}

In this section we discuss the behaviour of our non-relativistic fermion in the presence of
an AdS black hole, corresponding to finite temperature and finite chemical potential. One
may expect that as soon as the bulk theory is placed at finite temperature, all modes in
the flat band will become occupied. In fact, we will see that, at least to leading order in
$1/N$, this is not the case.

\para
The AdS Reissner-Nordstr\"om black hole metric is 
\be ds^2 = \frac{dr^2}{r^2f(r)} - r^2f(r)dt^2 + r^2 d\vec{x}^2\nn\ee
where the emblackening factor is given by
\be f(r) = 1 -\frac{1+Q^2}{r^3}+ \frac{Q^2}{r^4}\nn\ee
We have chosen to rescale the coordinates to place the black hole horizon at $r=1$ which means the temperature and chemical potential described below are both dimensionless. The background also contains an electric field, with gauge potential given by
\be A_0=gQ\left(1-\frac{1}{r}\right)\nn\ee
where the gauge coupling $1/g^2$ is the coefficient in front of the bulk Maxwell action. In the context of AdS/CFT, this gauge coupling arises in the current two-point function, $\langle JJ\rangle \sim 1/g^2$, where it
is loosely interpreted as the amount of charged stuff living among the $N^2$ strongly coupled degrees of freedom of the boundary field theory. In what follows, we will refer to this $N^2$ degrees of freedom as the soup.

\para
As usual, this black hole background places the strongly coupled soup at finite temperature $T$ and chemical potential $\mu$, given by
\be T = \frac{1}{4\pi}(3-Q^2)\ \ \ \ ,\ \ \ \ \mu=gQ\nn\ee
We probe this soup with our flat band fermion. The bulk action \eqn{bulkaction} remains the same, but the covariant derivative now conceals a coupling to the background gauge field $D_M = \partial_M + \ft14 \omega_{ab,M}\Gamma^{ab} - iqA_M$. We will set the charge $q=1$ and study how the physics of the probe changes as we vary the gauge coupling $g$. Note that, by a rescaling of the gauge field, this is entirely equivalent to varying $q$. However, it is somewhat simpler to interpret variations of the gauge coupling: as $g$ increases, the amount of charged matter sitting in the soup decreases and the probe fermion gains prominence. In what follows, we work exclusively with a fermion of mass $m=0$. 
For masses $m\neq 0$, it is less easy to separate the eigenvalues $\lambda_+$ and $\lambda_-$ for all values of $\vec{k}$ and $\omega$ but invariant objects like $ {\rm Tr}\, {\cal G}_R$ do not exhibit  qualitatively different  behaviour from the $m=0$ story we tell below.

\begin{figure}[t]
   \centering
    \begin{minipage}[b]{8 cm}
    \includegraphics[scale=0.45]{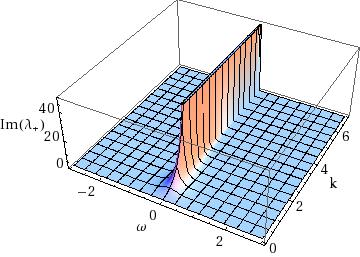}  
  \end{minipage}
  \begin{minipage}[b]{6 cm}
    \includegraphics[scale=0.45]{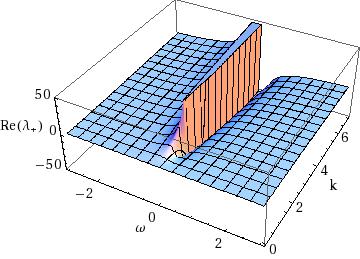}  
  \end{minipage}
  \caption{The imaginary (left) and real (right) parts of the flat band eigenvalue, $\lambda_+$, at finite temperature}
   \label{fignz}
\end{figure}

\subsection*{Finite Temperature}

We start by setting $g=0$, corresponding to placing the boundary field theory at
 finite temperature, but vanishing chemical potential. The spectral function for 
the singular, flat band eigenvalue $\lambda_+$ is shown in the left hand plot of 
Figure \ref{fignz}. Strikingly, although there is a depletion in the low momentum 
modes, the spectral function for the high momentum modes is essentially unchanged 
from the vacuum. In particular, there is an $\omega=0$ peak which becomes strongly 
delta functionesque as $\vec{k}$ increases as  evidenced by the pole in the real 
part of the Green's function which is plotted on the right of Figure \ref{fignz}. It is 
unusual to see such dramatic response at energies below the temperature and suggests 
that there are zero energy fermionic modes which have not been excited even though 
the system is at finite temperature. These modes persist because they sit outside the 
lightcone and cannot therefore decay.

\para
To understand physically what is happening here, we need to realise that the black hole background does not directly heat up the flat band. Rather, it heats up the CFT soup to a temperature which, in our convention, is $T=3/4\pi$. This hot soup is subsequently probed by the bulk fermion. Usually, one expects that the probe and soup equilibrate, so that all probe states up to 
energy $T$ are excited. Yet this does not  happen for the flat band, where $\omega=0$ 
states remain unoccupied. It would seem that the question of equilibrium is more subtle. 
Specifically, the soup is relativistic and has momentum modes excited up to  
$|\vec{k}| \sim T$. Since momentum is conserved, these soft modes can only excite 
the high momentum modes of the flat band under the unlikely condition that many 
collide at once. Thus, the spectral function of the flat band is determined not 
just by the temperature, but by the momentum distribution of the soup and its 
coupling to the fermion.

\begin{figure}[t]
   \centering
    \begin{minipage}[b]{8 cm}
    \includegraphics[scale=0.5]{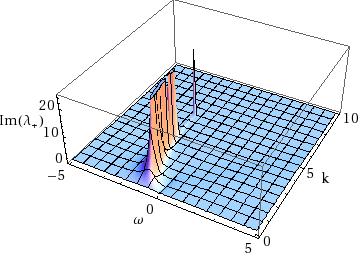}  
  \end{minipage}
  \begin{minipage}[b]{6 cm}
    \includegraphics[scale=0.5]{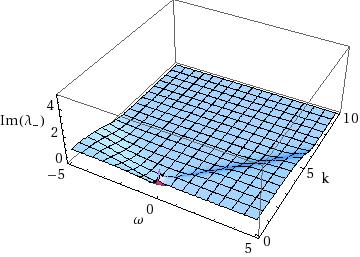}  
  \end{minipage}
  \caption{The imaginary parts of the eigenvalues  $\lambda_+$ (left) and $\lambda_-$ (right) for $g=1$}
   \label{figg1}
\end{figure}

\begin{figure}[t]
   \centering
    \begin{minipage}[b]{9 cm}
    \includegraphics[scale=0.5]{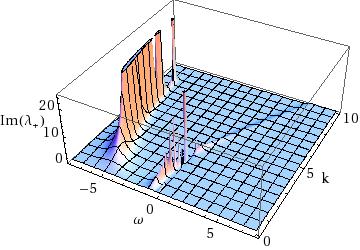}  
  \end{minipage}
  \begin{minipage}[b]{6 cm}
    \includegraphics[scale=0.35]{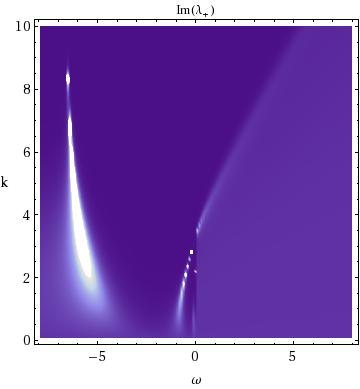}  
  \end{minipage}
  \caption{The imaginary part of the eigenvalue  $\lambda_+$ for $g=4$}
   \label{figg4}
\end{figure}

\begin{figure}[t]
   \centering
    \begin{minipage}[b]{9 cm}
    \includegraphics[scale=0.5]{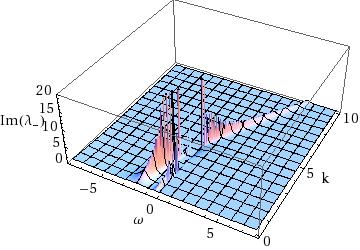}  
  \end{minipage}
  \begin{minipage}[b]{6 cm}
    \includegraphics[scale=0.35]{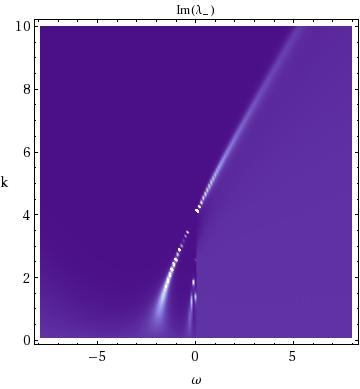}  
  \end{minipage}
  \caption{The imaginary parts of the eigenvalue $\lambda_-$  for $g=4$}
   \label{figbird}
\end{figure}

\subsection*{Finite Density}

We now turn on the chemical potential, focussing on $T=0$. In Figure \ref{figg1}, we 
plot both eigenvalues of the spectral function, ${\rm Im}\lambda_\pm$ for $g=1$. In 
the left-hand figure, ${\rm Im}\lambda_+$ clearly shows the flat band. Because 
frequency is now measured relative to the chemical potential, the band is shifted 
to $\omega = -\mu=-\sqrt{3}g$. Once again, we see that strong peaks remain at high momenta, 
reflecting the fact that these states are not occupied. (The apparent disappearance 
of the peak at high $\vec{k}$ is numerical artifact. It arises because the peak 
tends sharper as can be checked by plotting a suitable 
cross-section at fixed $\vec{k}$). This means that, despite equality in their energies, the 
flat band is occupied from the low momentum modes up and, for this reason, does not 
contribute to the ground state degeneracy in the background of the Reissner-Nordstr\"om black hole. For low momenta, the flat band becomes mildly dispersive, albeit with a large width. 
For $g=1$, the spectral function for $\lambda_-$ is essentially featureless.

\para
In Figures \ref{figg4} and \ref{figbird} we crank up the gauge coupling to $g=4$ and, to illustrate the features, present a birds eye view of the spectral function together with the landscape plots. Once again, the flat band is clearly visible in Figure \ref{figg4}. However, now both  eigenvalues $\lambda_+$  and $\lambda_-$  reveal  a Fermi surface. 
 The presence of the flat band does not seem to greatly alter the Fermi surface.   Indeed, it can be checked that the quasi-particle excitation spectrum (or lack thereof) is governed by the AdS${}_2$ near horizon regime of the black hole, as explained in \cite{mit2}. At most, the flat band appears to mildly suppress the formation of the Fermi surface. For example, note that in both  standard and alternative quantizations (which are effectively the same for $m=0$), the Fermi surfaces is already present at $g=1$. Moreover, as we increase $g$, the Fermi surface appears in the $\lambda_-$ eigenvalue before it appears in $\lambda_+$.

\begin{figure}[t]
   \centering
    \begin{minipage}[b]{8 cm}
    \includegraphics[scale=0.4]{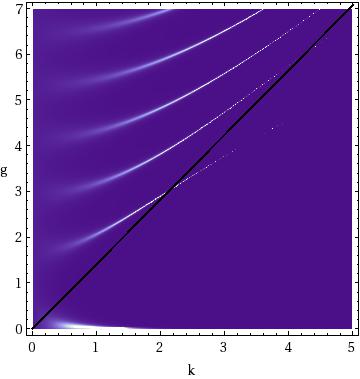}  
  \end{minipage}
  \begin{minipage}[b]{6 cm}
    \includegraphics[scale=0.4]{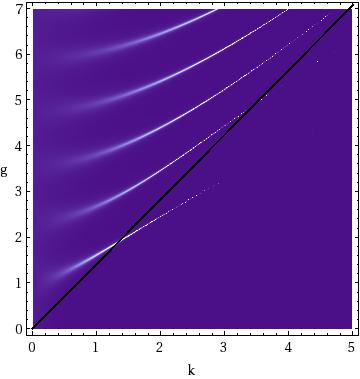}  
  \end{minipage}
  \caption{The density plot of ${\rm Im}\,\lambda_+$ (left) and 
${\rm Im}\,\lambda_-$ (right) as $g$ varies. These plots were made at $\omega = -10^{-2}$ and 
are robust to lowering the value of $|\omega|$.}
   \label{figpurple}
\end{figure}

\para
This latter point is highlighted in Figure \ref{figpurple} which depicts the 
$\omega=0$ density plot as $g$ is varied. 
 Above the black, diagonal line lies the oscillatory region, defined in 
\cite{mit2}, where the IR exponent is complex and no Fermi surface can form. 
The Fermi surfaces appear when the white lines extend below the black, where 
they are delta functions revealing themselves as faint, numerical pinpricks of 
paint. Note also the splash of white along the $g=0$ axis in the left-hand plot, 
corresponding to the flat band.

\begin{figure}[t]
   \centering
    \begin{minipage}[b]{8 cm}
    \includegraphics[scale=0.42]{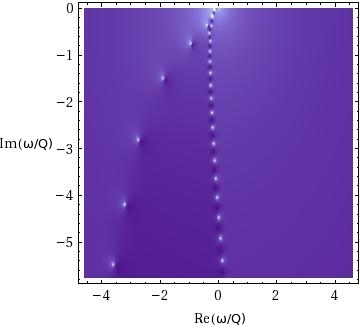}  
  \end{minipage}
  \begin{minipage}[b]{7 cm}
    \includegraphics[scale=0.42]{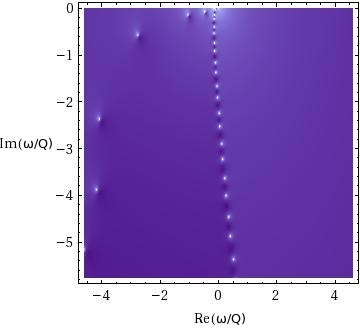}  
  \end{minipage}
   \begin{minipage}[b]{8 cm}
    \includegraphics[scale=0.42]{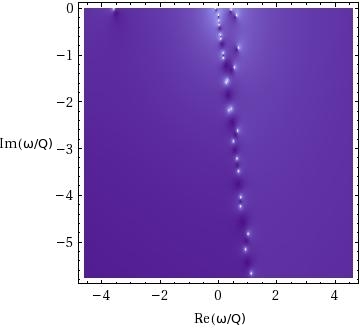}  
  \end{minipage}
   \begin{minipage}[b]{7 cm}
    \includegraphics[scale=0.42]{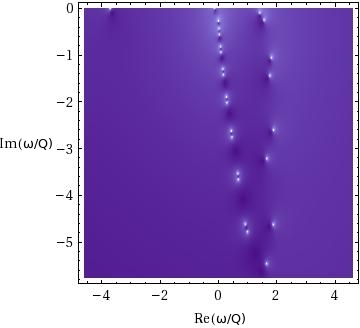}  
  \end{minipage}
  \caption{The quasi-normal mode spectrum for $g=4$ and $k=0.1$, $k=0.3$, $k=5$ and $k=7$}
   \label{qnm}
\end{figure}

\para
Finally, in Figure \ref{qnm} we plot the quasi-normal mode spectrum for $T=0$, $g=4$ and increasing values of the momentum. The plots show $k=0.1$, $k=0.3$, $k=5$ and $k=7$ respectively. They were computed using an extension of the scalar method presented in \cite{leaver}. (See also \cite{Denef:2009kn}). Here one Taylor expands the Dirac equation in the black hole background, transforming the differential equation into a matrix equation. This expansion is truncated at finite order, resulting in a large but finite matrix and, for fixed $\vec{k}$, one scans the complex $\omega$ plane for points where the determinant  vanishes. These are the quasi-normal modes, plotted for both $\lambda_+$ and $\lambda_-$ in the diagrams above. (These diagrams were made after truncating to a $200\times 200$ matrix). As the quasi-normal modes approach the real axis, they coincide with peaks in the spectral function.

\para
The first feature in these plots is the ridge of quasi-normal modes running almost vertically downwards. These are the remnant of a branch cut. (Indeed, as we go to higher order, we more points appear in this ridge). For small $|\vec{k}|$, a number of quasi-normal modes fan out from this ridge in the  ${\rm Re}\omega <0$ direction. These are clearly visible in the first and second plots above. As $|\vec{k}|$ increases yet further, a number of these quasi-normal modes start to approach the real $\omega$ axis. For the  $g=4$ case shown, the first two quasi-normal modes cross over to ${\rm Re}\omega >0$ where they hit the real axis before bouncing off. These correspond to the Fermi surfaces shown in \ref{figg4} and \ref{figbird}. The first quasi-normal mode to hit is the $\lambda_+$ eigenvalue; the second is $\lambda_-$. Indeed, it can be verified that $\lambda_+$ exhibits a Fermi surface at lower $k_F$ which indeed exhibits a Fermi surface at lower $|\vec{k}|$ than $\lambda_-$. The third quasi-normal mode does something different: it too heads towards the real $\omega$ axis, but it doesn't bounce. It can be seen in the last two plots above, nudging closer to $\omega/Q = -g = -4$ where it remains for ever higher $\vec{k}$. This is the flat band.

\section*{Acknowledgement}
Our thanks to Sean Hartnoll, John McGreevy, Andy O'Bannon, Andrew Green, Miguel Paulos, Subir Sachdev,  D.T. Son, Julian Sonner, Andrei Starinets and Stefan Vandoren for useful discussions. JNL is supported by the Funda\c{c}$\tilde{\rm a}$o para a Ci$\hat{\rm e}$ncia e Tecnologia (FCT-Portugal) through the grant SFRH/BD/36290/2007 and partially supported by CERN/FP/116377/2010. DT is supported by the Royal Society and ERC STG grant 279943.

\end{document}